\documentclass[aps,twocolumn,amsmath,amssymb]{revtex4}
\usepackage{graphicx}
\usepackage[T1]{fontenc}
\usepackage{hyperref}
\hypersetup{
  colorlinks   = true,  
  urlcolor     = blue,  
  linkcolor    = blue,  
  citecolor    = red    
}
\begin{document}

\title{Few-fermion thermometry }
\author{Marcin P{\l}odzie\'n$^1$, Rafa{\l} Demkowicz-Dobrza\'nski$^2$, and Tomasz Sowi\'nski$^1$}
\affiliation{
\mbox{$^1$ Institute of Physics, Polish Academy of Sciences, Aleja Lotnikow 32/46, PL-02668 Warsaw, Poland}
\mbox{$^2$ Faculty of Physics, University of Warsaw, Pasteura 5, PL-02093 Warsaw, Poland}
}

\begin{abstract}
Potential realization of a quantum thermometer operating in the nanokelvin regime, formed by a few-fermionic mixture confined in a one-dimensional harmonic trap, is proposed. Thermal states of the system are studied theoretically from the point of view of fundamental sensitivity to temperature changes. It is pointed out that the ability to control the interaction strength in such systems allows obtaining high-temperature sensitivity in the regime where the temperature is much lower than the characteristic temperature scale determined by a harmonic confinement. This sensitivity is very close to the fundamental bound that involves optimal engineering of level separations. The performance of practical measurement schemes and the possible experimental coupling of the thermometer to the probe are discussed.
\end{abstract}

\maketitle

\section{Introduction}
The variety and sophistication of present-day temperature measurement techniques are stunning \cite{Childs2000ReviewTemp}.
While most practical techniques rely on the quantum features of matter or light in an indirect way, the recent rapid development of
quantum technology related to experimental techniques opened both new possibilities and challenges to temperature measurements.
Among others, sub-millikelvin thermometry techniques have been developed
for the purpose of assessing the efficiency of cooling in ion trap experiments \cite{Knunz2012IonThermetry, Norton2011SpatialThermometry},
 properties of nitrogen-vacancy centers have been utilized for high sensitivity all-optical thermometry with potential application in biomedical research \cite{Kucsko2013, Plakhotnik2014}, and new techniques to measure a temperature of quantum dots \cite{Braun2013,Seilmeier2014,Haupt2014,Chekhovich2017},
 cold degenerate quantum gases \cite{Lous2017}, optomechanical systems \cite{Brunelli2011}, atoms in optical lattices \cite{Roustekoski2009},  or quantum impurity in Bose-Einstein condensates \cite{Sabin2014} have been proposed. In parallel, there has been a growing
 interest in understanding theoretical limits to temperature estimation precision
 from a fundamental point of view of quantum estimation theory \cite{Stace2010, Monras2011, Brunelli2012, Correa2015, Jarzyna2015, Paris2016, Mukherjee2017, Hofer2017,Correa2017,Correa2017b,Campbell2018}.
 Most prominently, in \cite{Correa2015} the optimal level structure for an $M$-level thermalized quantum system was identified that leads to the highest sensitivity to temperature changes. The structure consists of a single non-degenerate ground state and an $M-1$ fold degenerate excited state with an energy gap proportional to the temperature. It turns out that an energy structure with $M$ isolated levels is naturally present in the systems of two-component fermionic mixtures confined in a one-dimensional harmonic traps \cite{Sowinski2013,Gharashi2013}. Although the level structure is different from the optimal one predicted in \cite{Correa2015}, these kinds of systems can be deterministically prepared, precisely controlled, and deeply analyzed in nowadays experiments \cite{serwane2011deterministic,wenz2013fewToMany,zurn2012fermionization,zurn2013Pairing}. In consequence, a brand new path of exploration of few-body problems in the context of ultra-cold atoms is opened (for review see for example \cite{Blume2012Rev,Zinner2016Rev}).

Motivated by these observations in this paper, we study temperature sensitivity of a \textit{fully thermalized} system consisting of a few interacting fermions in a harmonic trap, already thermalized with the probe. In case of noninteracting particles and in the very low-temperature limit, where $k_BT \ll \hbar \Omega$ ($\Omega$ being the harmonic trap frequency) small variations of temperature will not affect the harmonic trap populations. However, the possibility to tune the interaction strength in such systems allows changing the energy level structure in such a way that the lowest energy states become almost degenerate with a well-controlled energy gap. We show that this system manifests temperature sensitivity that approaches surprisingly close the fundamental theoretical bound from \cite{Correa2015}. Furthermore, while optimal measurement extracting the full temperature information may not be practical, we show that simple single-particle population measurements still allow reaching high-temperature sensitivities surpassing the non-interacting reference case by many orders of magnitude.

The paper is organized as follows. In section~II we comment on fundamental bound for the accuracy of the temperature measurements in the language of the Quantum Fisher Information (QFI). Next, in section~III we introduce an experimentally realizable \textit{quantum thermometer} formed by an ultra-cold two-component mixture of a few fermions confined in a harmonic trap and we compare its QFI with the optimal system considered in \cite{Correa2015}. In section~IV we discuss three different protocols indicating an experimentally accessible measurement. We also deliberate on a possible coupling of the thermometer to a probe of unknown temperature. Finally, we conclude in section V.

\section{The fundamental bound}
Let us consider an $M$-level quantum system prepared in a thermal state described by the  density matrix
\begin{equation}
\hat\rho_T = \frac{1}{\cal Z}{\mathrm{e}^{-\beta\hat{\cal H}}}=\frac{1}{\cal Z}\sum_{i=1}^M \mathrm{e}^{-\beta{\cal E}_i}|\mathcal{E}_i\rangle\langle \mathcal{E}_i|,
\end{equation}
where  $|{\mathcal{E}_i}\rangle$ are eigenstates of the system Hamiltonian $\hat{\cal H}$ with respective energies $\mathcal{E}_i$, $\beta=1/k_B T$, and the partition function ${\cal Z}=\sum_i \mathrm{e}^{-\beta{\cal E}_i}$.
The maximal information on temperature variations that can be extracted from this state is quantified with the Quantum Fisher Information (QFI) \cite{Helstrom1976, Braunstein1994}:
\begin{equation}
{\cal F}_Q(\hat\rho_T) = 4\sum_{m,n}p_m\frac{|\langle {\cal E}_m|\partial_T\hat{\rho}_T |{\cal E}_n\rangle|^2}{(p_m+p_n)^2} = \frac{\varDelta{\cal H}^2}{T^4},
\end{equation}
where $p_i=\langle\mathcal{E}_i| \hat{\rho}_T |\mathcal{E}_i\rangle={\cal Z}^{-1}\mathrm{e}^{-\beta\mathcal{E}_i}$ and $\varDelta {\cal H}^2 = \text{Tr}[\hat{\rho}_T \hat{\cal{H}}^2] - \text{Tr}[\hat{\rho}_T \hat{\cal H}]^2$ is the variance of the thermal expectation value of the system Hamiltonian \cite{Jing2014}.
In the energy eigenbasis this coincides with the classical Fisher Information (FI) for the state energy probability distribution
\begin{equation}
{\cal F}_Q(\hat\rho_T) = {\cal F}(\{p_i\})=\sum_i \frac{1}{p_i} \left(\frac{\mathrm{d}p_i}{\mathrm{d}T}\right)^2.
\end{equation}
Given $\nu$ repetitions of an experiment, the QFI yields a fundamental lower bound on the minimal temperature estimation precision via the
quantum Cram{\'e}r-Rao inequality
\begin{equation}
\Delta T \geq \frac{1}{\sqrt{\nu {\cal F}_Q}}.
\end{equation}
The bound can always be saturated in the $\nu \rightarrow \infty$ limit, and in the considered case the optimal measurement is the  measurement in the energy eigenbasis.

As discussed in \cite{Correa2015}, for a given temperature $T$, the optimally engineered level structure in a $M$-level system leads to the following QFI and the respective temperature sensitivity bound  (we write it here in a bit more explicit way than in the original paper):
\begin{equation}
\label{eq:qfimax}
{\cal F}^{\mathrm{max}}_Q(T) = \frac{f(M)}{T^2},\quad   \frac{\Delta T}{T} \geq \frac{1}{\sqrt{f(M)}},
\end{equation}
where $f(M) = (M-1)(M-1+e^x)^{-2}e^x x^2$, while $x\geq 0$ is the solution of the following transcendental equation
$e^x = (M-1)(2+x)/(2-x)$. To get an intuition on the above result, we inspect the $M\rightarrow \infty$ limit in which case
$x \approx \ln M$, and $f(M) \approx (\ln M)^2/4$. Consequently, $\Delta T/T \geq 2/\ln M$ reflects a logarithmic reduction of the relative uncertainty of temperature measurements with the increasing size of the system. The optimal level structure leading to the above QFI involves a single ground state and an $M-1$ degenerate excited state with the optimal energy gap equal to $\mathcal{E}_2 - \mathcal{E}_1  = x\,k_BT$ \cite{Correa2015}.

Eq.~\eqref{eq:qfimax} can be regarded as the fundamental bound for thermometry utilizing thermalized quantum states. Note, however, that there are other approaches to thermometry that assume non-thermalized thermometer scenarios, such as e.g. temperature influencing indirectly the phase of light traveling through an interferometer \cite{Stace2010, Jarzyna2015} or temperature affecting the dissipative character of quantum state evolution  \cite{Monras2011}. In such scenarios, one will arrive at different model-specific bounds for temperature estimation which often involve non-trivial optimization over a large class of input probe states. In case of thermometry utilizing thermalized quantum states that we consider here, the only mean of adjusting the probe state is via modification of the level structure of the system. While this may be viewed as a deficiency of the approach, as we do not benefit from the typical quantum metrological enhancements offered by entangled or squeezed states, the advantage here is that the considerations are model-independent and they are based on a simple and natural physical assumption of thermalization.

\section{The system}
\begin{figure}
\includegraphics[scale=0.8]{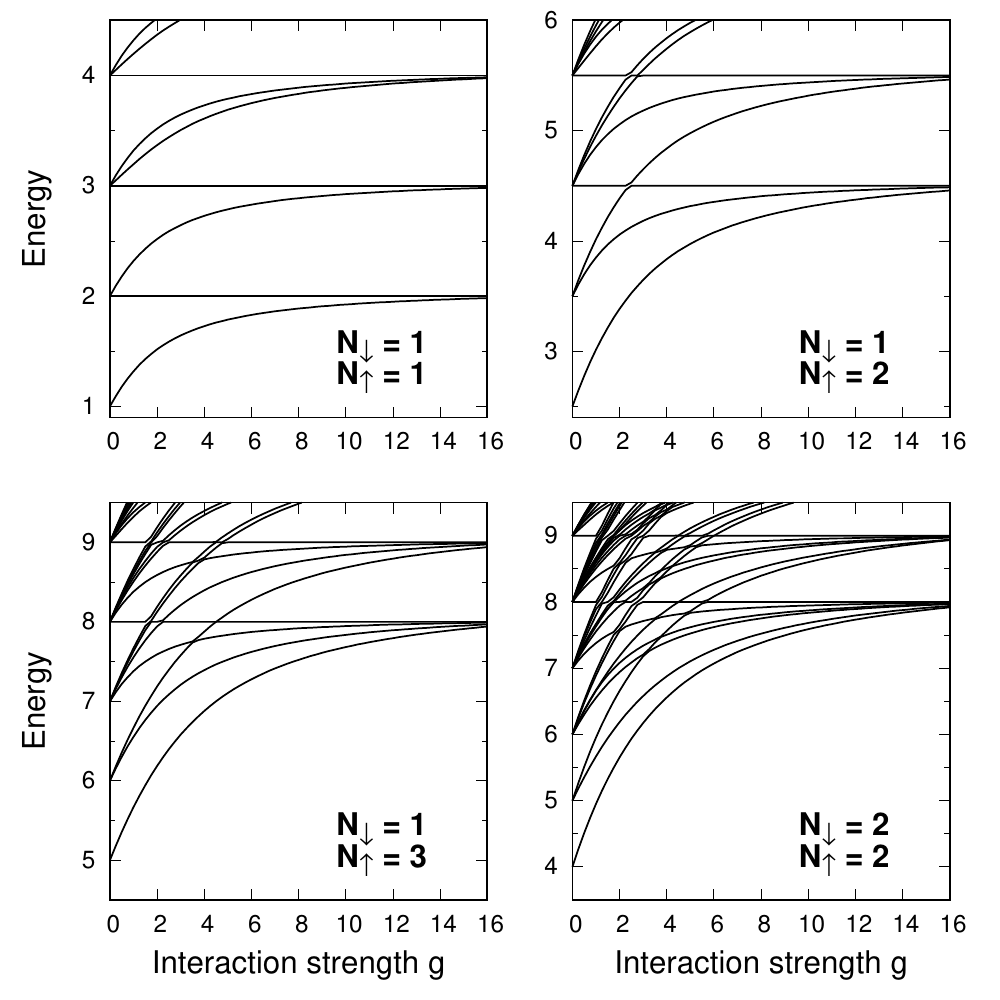}
\caption{Energy spectrum of the many-body Hamiltonian \eqref{HamFerm} for different number of particles as a function of interaction strength $g$. Note the quasi-degeneracy of the many-body states for strong repulsions. The energies and the interaction strengths are measured in units of $\hbar\Omega$ and $(\hbar^3\Omega/m)^{1/2}$, respectively.\label{Fig1}}
\end{figure}
The considered energy level structure with quasi-degeneracy is naturally present in the system of interacting two-component mixture of a few ultra-cold fermions confined in a one-dimensional harmonic trap, where Hamiltonian reads \cite{Sowinski2013,Gharashi2013}:
\begin{align} \label{HamFerm}
\hat{\cal H} &= \sum_\sigma\int\!\!\mathrm{d}x\,\hat\Psi_\sigma^\dagger(x)\left(-\frac{\hbar^2}{2m}\frac{\mathrm{d}^2}{\mathrm{d}x^2}+\frac{m\Omega^2}{2}x^2\right)\hat{\Psi}_\sigma(x) \nonumber \\
&+g\int\!\!\mathrm{d}x\,\hat\Psi_\downarrow^\dagger(x)\hat\Psi_\uparrow^\dagger(x)\hat{\Psi}_\uparrow(x)\hat{\Psi}_\downarrow(x).
\end{align}
Here, $\hat{\Psi}_\sigma(x)$ is a fermionic field operator corresponding to the component $\sigma\in\{\uparrow,\downarrow\}$ and obeying anti-commutation relations, $\{\hat\Psi_\sigma(x),\hat\Psi_{\sigma'}^\dagger(x')\}=\delta_{\sigma\sigma'}\delta(x-x')$ and $\{\hat\Psi_\sigma(x),\hat\Psi_{\sigma'}(x')\}=0$. Interaction strength $g$ is an effective parameter related to the three-dimensional s-wave scattering length between atoms and it can be tuned experimentally with a huge accuracy almost on demand \cite{Olshanii1998,Feshbach2010,haller2009realization,serwane2011deterministic,wenz2013fewToMany}. The Hamiltonian \eqref{HamFerm} commutes with the operators counting numbers of particles in a given spin $\hat{N}_\sigma = \int\mathrm{d}x\,\hat\Psi_\sigma^\dagger(x)\hat\Psi_\sigma(x)$.  Consequently, a whole analysis can be performed in subspaces of given number of particles in individual components  $N=N_\uparrow+N_\downarrow$. It is convenient to perform an analysis in the basis of single-particle orbitals $\varphi_i(x)$ being eigenstates of the corresponding single-particle Hamiltonian. In this basis the many-body Hilbert space is spanned by Fock states constructed as
\begin{multline}\label{Fock_state}
|\mathrm{F}_k\rangle \equiv |n_1, n_2, \ldots;m_1, m_2, \ldots\rangle \\ \sim (\hat{a}_{\uparrow 1}^\dagger)^{n_1}(\hat{a}_{\uparrow 2}^\dagger)^{n_2}\cdots(\hat{a}_{\downarrow 1}^\dagger)^{m_1}(\hat{a}_{\downarrow 2}^\dagger)^{m_2}\cdots|\mathtt{vac}\rangle,
\end{multline}
where operator $\hat{a}_{\sigma i}$ annihilates a particle with spin $\sigma$ in a state $\varphi_i(x)$. Due to the fermionic statistics and conserved numbers of particles following constrains have to be applied $n_i,m_i\in\{0,1\}$, $\sum_i n_i = N_\uparrow$ and $\sum_i m_i = N_\downarrow$.
We perform numerically exact diagonalization of the Hamiltonian \eqref{HamFerm} in the Fock basis $\{|\mathrm{F}_k\rangle\}$ appropriately cropped to states with the lowest energies \cite{Haugset1998,Plodzien2018}. In this way we obtain the lowest many-body eigenstates $|\mathcal{E}_i\rangle$ and corresponding eigenenergies ${\cal E}_i$ (see Fig.~\ref{Fig1}). The most pronounced feature of the spectrum is the quasi-degeneracy of the ground-state manifold in the regime of strong interactions, with degeneracy $M = \frac{N!}{N_\downarrow !\,N_\uparrow !}$ \cite{Guan2009,Sowinski2013}.
Intuitively, the larger degeneracy the higher sensitivities can be expected, but only in the regime where $k_BT$ is comparable with the energy width of the degenerated manifold corresponding to a given $g$. For a given number of particles $N$ the largest degeneracy is to be expected for balanced partition of particles into different spin components, $N_\downarrow = N_\uparrow = N/2$, and in the following, we focus on this configuration with $N=4$. Qualitatively, the behavior for other cases will be analogous, but some appropriate rescaling of the QFI needs to be performed, according to the change in the number of quasi-degenerate states.

\begin{figure}
\includegraphics[scale=0.8]{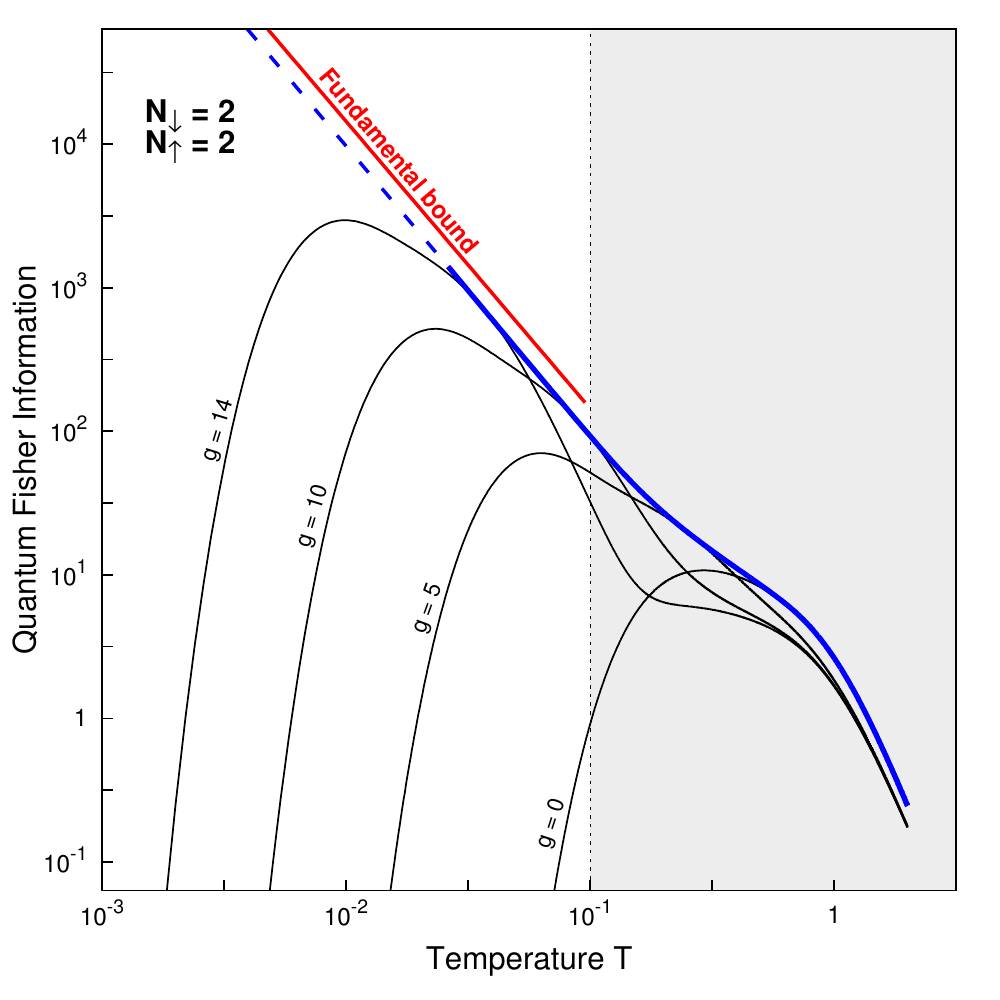}
\caption{Quantum Fisher Information for a balanced system of $N=4$ fermions, as a function of temperature $T$, for different interaction strengths $g$. The blue envelope corresponds to optimization of $g$ for each value of $T$ and approaches very closely the fundamental bound \eqref{eq:qfimax} for $M=6$ which is the quasi-degeneracy of the ground state. For $T \geq 0.1$ (shaded area), when
$k_BT$ becomes comparable with $\hbar \Omega$, higher energy levels start to contribute and hence the envelope may surpass the fundamental bound where only $M=6$ levels are considered. The QFI and temperature are measured in units of $(\hbar\Omega/k_B)^{-1/2}$ and $\hbar\Omega/k_B$, respectively.
 \label{Fig2}}
\end{figure}

\section{Measurement protocols}
In order to get an understanding of the maximum potential temperature sensitivity of the system,  in Fig.~\ref{Fig2} we plot QFI as a function of temperature for different $g$. We also provide the envelope curve (solid blue) which corresponds to the ultimate limit where $g$ is optimally chosen for a given temperature in order to assure the energy gap is optimal. Additionally, on top of this, we plot the fundamental bound given by \eqref{eq:qfimax} for $M=6$ states chosen so that it equals the quasi-degeneracy of the ground manifold of our system in strong repulsion regime, $g\rightarrow\infty$. The envelope obtained for the few-fermion system is always below the fundamental bound. This is a direct consequence of the structure of the energy spectrum which in the case studied is different from the optimal structure discussed in \cite{Correa2015}. The energy levels form a degenerate manifold rather than the gap structure between the isolated state and $M-1$ degenerated states. However, the discrepancy between the fundamental bound and the envelope is less than 40\% for any temperature. This is a surprisingly good result taking into account that we consider a naturally appearing level structure with only a single tuning parameter $g$.
\begin{figure}
\includegraphics[scale=0.8]{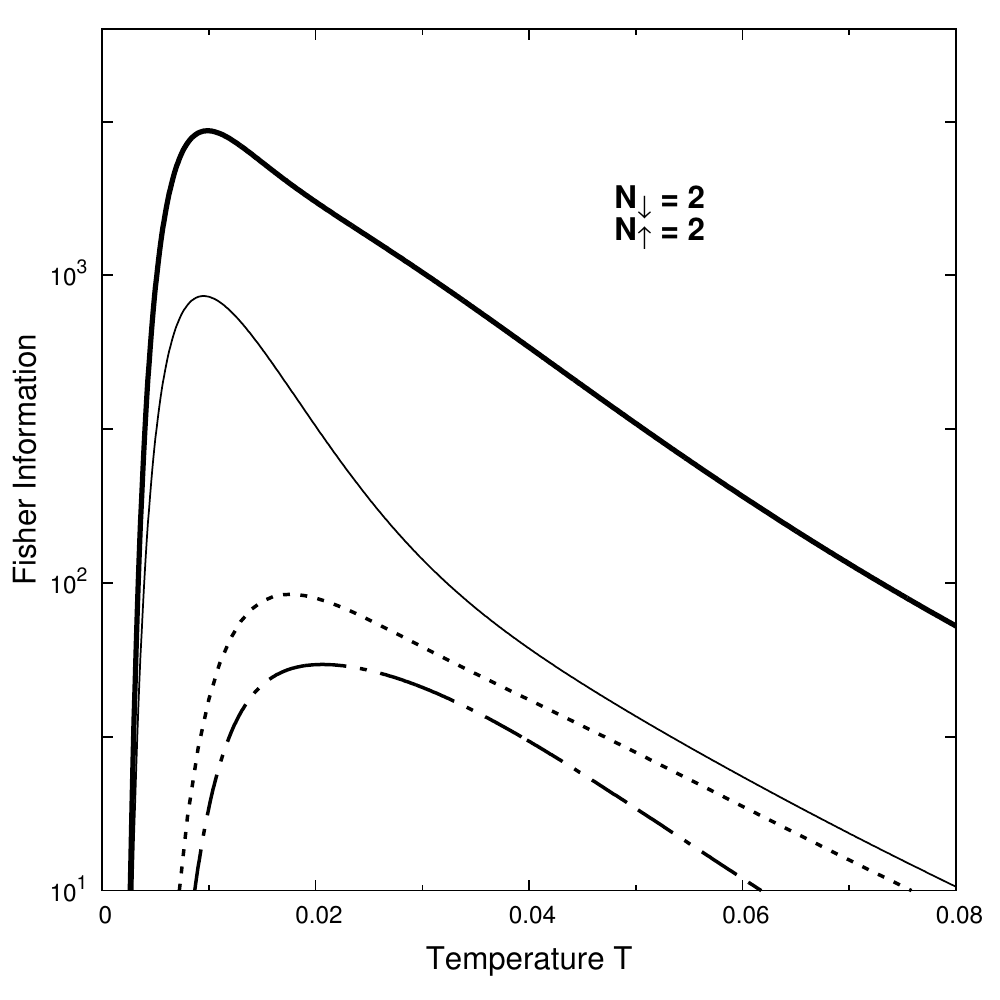}
\caption{Different Fisher informations as a function of temperature $T$ calculated for strongly interacting system ($g=14$) of $N_\uparrow=N_\downarrow=2$. The solid thick line corresponds to the Quantum Fisher information ${\cal F}_Q(T)$ which defines an upper limit for other Fisher information obtained via particular measurement schemes. Other lines represent different variants of Fock basis measurement: ${\cal F}_{\mathtt{Fock}}(T)$ -- complete Fock states projection  (solid thin line), ${\cal F}_{\mathtt{Coarse}}(T)$ -- measurement of total occupation of the two lowest single-particle orbitals (dotted line), ${\cal F}_{\mathtt{Fermi}}(T)$ -- binary measurement of presence of a particle just above the Fermi level (dotted-dashed line). See the main text for details. The Fisher information and temperature are measured in units of $(\hbar\Omega/k_B)^{-1/2}$ and $\hbar\Omega/k_B$, respectively. \label{Fig3}}
\end{figure}

Naturally, reaching the sensitivity predicted by the QFI requires measurements projecting the system onto energy eigenbasis and it may be very difficult in practice. From the experimental perspective, a much more feasible option is particle population measurement on respective single-particle orbitals. For the considered system the most general basis of this type is spanned by the Fock states of two-component mixture \eqref{Fock_state}. In the following, we consider three types of measurements prepared in the Fock basis. Each of these measurements depends on the \textit{resolution} available in the many-body basis. Next, in Fig.~\ref{Fig3} we compare the Fisher information obtained for these different measurement schemes and with the QFI encoding the impassable bound for given experimental realization.

The most general measurement in the Fock basis \eqref{Fock_state} is based on simple projections of the many-body thermal state $\hat\rho_T$ on a single many-body Fock state $|\mathrm{F}_i\rangle$, {\it i.e.}, $p_i(T) = \mathrm{Tr}(\hat\rho_T\,|\mathrm{F}_i\rangle\langle\mathrm{F}_i|)$. Corresponding FI, ${\cal F}_{\mathtt{Fock}}(T)$, is shown in Fig.~\ref{Fig3} with a solid thin line. As suspected, some reduction of the temperature sensitivity is present. Although this approach is very general, it requires many accurate and demanding measurements of occupations on all possible single-particle levels. Therefore, it is more reasonable to assume that one has only limited access to many-body Fock states and rather it is only possible to perform measurements of coarse probabilities on some low-lying single-particle orbitals. In the considered case of four fermions, the simplest measurement of this kind gives as an output the probability $P_k$ of finding exactly $k = 0,\ldots,4$ particles occupying two the lowest orbitals of the harmonic trap \cite{serwane2011deterministic}. This kind of measurements correspond to coarse-grained projectors $\mathcal{P}_k = \sum_i |\mathrm{F}_i^k\rangle\langle\mathrm{F}_i^k|$, where summation runs over all Fock states having exactly $k$ particles in two the lowest orbitals. Simply, the probabilities $P_k$ are calculated straightforwardly by dividing the Fock basis $\{|\mathrm{F}_i\rangle\}$ to five independent subsets having exactly $k$ particles in chosen orbitals. We denote the corresponding FI as ${\cal F}_{\mathtt{Coarse}}(T)$ and we plot it in Fig.~\ref{Fig3} with the dotted line. Obviously, the coarse-grained measurements significantly reduce the number of possible outcomes from the dimension of considered Hilbert space to only five numbers. Therefore, reduction of FI is suspected. Surprisingly, as seen in Fig.~\ref{Fig3}, this reduction is only by one order of magnitude when compared to measurements based on individual Fock states. These results show that experimentally accessible measurements may serve as appropriate and relevant tools from the metrological point of view.

Finally, we also analyze the simplest measurement from the experimental point of view. It is based on a direct measurement of the probability of finding a particle with a given spin $\sigma$ on a first excited state above the Fermi sea of the non-interacting system. The measurement results in a binary outcome $\{P, 1-P\}$ related to a single-particle number operator $\hat{n}_{\sigma,i_F}=\hat{a}^\dagger_{\sigma,i_F}\hat{a}_{\sigma,i_F}$, where $i_F$ is an index of a first single-particle orbital above the Fermi level of non-interacting system (in the case of $N_\uparrow=N_\downarrow=2$ particles $i_F=3$). The corresponding FI, ${\cal F}_{\mathtt{Fermi}}(T)$, is displayed with the dotted-dashed line in Fig.~\ref{Fig3}. As it is seen, the FI obtained for a binary outcome is of the same order of magnitude as the FI obtained via coarse-grained Fock space measurements and is only two orders of magnitude smaller when compared to the optimal QFI. Although the information obtained with available experiments is reduced when simpler measurement schemes are considered, even the elementary binary measurement provides the sensitivity that is three orders of magnitude higher than in the non-interacting case. As such, a considered system forms a relatively good \textit{quantum thermometer} operating in the regime of tens of nK (for standard trapping frequencies operating on the order of kHz \cite{serwane2011deterministic}).

The quantum thermometer can be practically utilized only if there exists a route for weak coupling and thermalization with some other quantum system (in this case some other ultra-cold gas) serving as a probe of unknown temperature. In the considered case such a coupling is indeed possible  and can be carried out by any long-range interaction leading to the unconstrained energy transfer between systems without introducing significant correlations. One of the possible paths is to exploit the Rydberg-dressing technique \cite{Johnson2010,Honner2010}, {\it i.e.}, the off-resonant coupling of the probe and thermometer to two different highly lying Rydberg states with different orbital quantum numbers. Although, symmetric dressing results only in overall static energy shift in both systems \cite{Balewski2014,Plodzien2017}, an unsymmetric off-resonant coupling to different Rydberg states results in weak Rydberg-dressed dipole-dipole interactions \cite{Wuster2011}. In consequence, a direct exchange of angular momentum and energy is present and may lead to the thermalization. Nevertheless, it should be pointed that the general problem of the thermalization in quasi-one-dimensional systems is still a long-standing goal and it is far beyond a scope of this work \cite{Alessio2016,Rigol2008,Kaminishi2015,Kaufman2016}.

\section{Conclusions}
Based on a parameter estimation theory we analyzed a thermal sensitivity of a two-component mixture of ultra-cold fermions confined in a one-dimensional harmonic trap in terms of the Quantum Fisher Information. We showed that the natural quasi-degeneracy of the many-body spectrum puts the system as a good candidate for an experimental realization of the quantum thermometer which was recently proposed theoretically by Correa {\it et al.} \cite{Correa2015}. An essential advantage of the considered setup is its potential tunability in the regime of temperatures on which the thermometer operates. The sensitivity of the system can be controlled straightforwardly by tuning inter-particle interactions and the trapping frequency. Controllability is increased since the number of states forming a quasi-degenerate manifold can be engineered by changing the number of particles. In principle, the system can be deterministically coupled to other quantum systems via the Rydberg dressing mechanism which allows performing experimental validation of the system. As shown, the Fisher information predicted for experimentally accessible measurements is only two orders of magnitude smaller than the fundamental bound determined by the Quantum Fisher Information.

\section*{Acknowledgements}
This work was supported by the (Polish) National Science Center Grants No. 2016/22/E/ST2/00555 (MP, TS) and 2016/22/E/ST2/00559 (RDD). We thank Micha{\l} Tomza for fruitful discussions.

\bibliography{_bibtex}
\end{document}